\documentclass[11pt]{article}
\usepackage{graphicx}
\usepackage[margin=1.25in]{geometry}
\usepackage[usenames,dvipsnames]{color}
\usepackage{url}

\usepackage{subfigure} 
\usepackage{bm}        
\usepackage{amsmath}   
\usepackage{cancel} 
\usepackage{verbatim} 

\usepackage[colorlinks = true,
            linkcolor = blue,
            urlcolor  = blue,
            citecolor = blue,
            anchorcolor = blue]{hyperref}

\newcommand\pubnumber{FERMILAB-FN-1146-AD}
\newcommand\pubdate{\today}

\def\Title#1{\begin{center} {\LARGE #1 } \end{center}}
\def\Author#1{\begin{center}{ \sc #1} \end{center}}
\def\Address#1{\begin{center}{ \it #1} \end{center}}

\newcommand\pubblock{\rightline{\begin{tabular}{l} \pubnumber\\
         \pubdate \end{tabular}}}
\newenvironment{Abstract}{\begin{quotation} \begin{center}
                       ABSTRACT
     \end{center}\bigskip  }{\end{quotation}}

\def\beq{\begin{equation}}
\def\eeq#1{\label{#1}\end{equation}}
\def\eeqn{\end{equation}}

\newenvironment{Eqnarray}%
   {\arraycolsep 0.14em\begin{eqnarray}}{\end{eqnarray}}
\def\beqa{\begin{Eqnarray}}
\def\eeqa#1{\label{#1}\end{Eqnarray}}
\def\eeqan{\end{Eqnarray}}

\let\bar=\overbar

\def\lsim{\mathrel{\raise.3ex\hbox{$<$\kern-.75em\lower1ex\hbox{$\sim$}}}}
\def\gsim{\mathrel{\raise.3ex\hbox{$>$\kern-.75em\lower1ex\hbox{$\sim$}}}}

\def\del{\partial}
\def\Dslash{\not{\hbox{\kern-4pt $D$}}}
\def\dslash{\not{\hbox{\kern-2pt $\del$}}}
\def\pslash{\not{\hbox{\kern-2pt $p$}}}
\def\ETmiss{\not{\hbox{\kern-4pt $E$}}_T}

\def\Dlr{\mathrel{\raise1.5ex\hbox{$\leftrightarrow$\kern-1em\lower1.5ex\hbox{$D$}}}}

\def\MSB{{\bar{M \kern -2pt S}}}
\def\msb{{\bar{\scriptsize M \kern -1pt S}}}

\def\drb{{\bar{\scriptsize D \kern -1pt R}}}

\newcommand\snowmass{\begin{center}\rule[-0.2in]{\hsize}{0.01in}\\\rule{\hsize}{0.01in}\\
\vskip 0.1in Submitted to the  Proceedings of the US Community Study\\ 
on the Future of Particle Physics (Snowmass 2021)\\ 
\rule{\hsize}{0.01in}\\\rule[+0.2in]{\hsize}{0.01in} \end{center}}

\begin{document}

\pubblock

\Title{Design Considerations for Fermilab Multi-MW Proton Facility in the DUNE/LBNF era}

\bigskip 
\Author{J.~Eldred, S.~Nagaitsev{\footnote{also at the University of Chicago}}, V.~Shiltsev, A.~Valishev, R.~Zwaska}

\Address{Fermi National Accelerator Laborator, Batavia, IL, USA}

\Author{M.~Syphers}

\Address{Northern Illinois University, Dekalb, IL, USA}

\medskip

 \begin{Abstract}
\noindent Fermilab has submitted two Snowmass whitepapers on a future 2.4~MW upgrade for DUNE/LBNF featuring a 2~GeV extension of the PIP-II linac and the construction of a new rapid-cycling-synchrotron. This paper summarizes the relationship between these two scenarios, emphasizing the commonalities and tracing the differences to their original design questions. In addition to a high-level summary of the two 2.4~MW upgrade scenarios, there is a brief discussion of staging, beamline capabilities, subsequent upgrades, and relevant R\&D. We are proposing a vigorous program to address various challenges associated with each scenario and to down-select the concept, most suitable to provide proton beams for years to come.
\end{Abstract}

\snowmass

\def\thefootnote{\fnsymbol{footnote}}
\setcounter{footnote}{0}

\section*{Executive Summary}

DUNE/LBNF constitutes an international multi-decadal physics program for leading-edge neutrino science and proton decay studies~\cite{DUNE} and is expected to serve as the flagship particle experiment based at Fermilab.

The Fermilab Main Injector (MI) is expected to provide 1.2~MW at 120~GeV for the DUNE/LBNF program, with the PIP-II upgrade~\cite{PIP2}. However the DUNE/LBNF science program also anticipates an upgrade of the Fermilab proton complex to 2.4~MW at 120~GeV in the Main Injector. The upgrade to 2.4~MW beam power for a 120-GeV MI cycle should also enable at least 2.15~MW for a 80-GeV MI cycle and 2.0~MW for a 60-GeV MI cycle.

There are two Snowmass white papers describing well-developed scenarios for a 2.4~MW upgrade of the Main Injector by extending the PIP-II linac energy (to 2~GeV) and constructing a new rapid-cycling synchrotron (RCS). The two white papers are what this document will call the modern Initial Configuration Document 2 (ICD-2) scenario~\cite{Nagaitsev} and the more expansive Booster science replacement (BSR) scenario~\cite{SyphersEldred}. This document will summarize the two scenarios, compare the design features, and indicate key planning considerations. There is a number of areas where Snowmass planning for particle physics experiments and accelerator technology could be instructive. The objective is to compare these scenarios and to propose the R\&D program, which is essential for the down-selection process.

Other than technical considerations and (likely) overall cost, the RCS scenarios differ from each other primarily in the beam power available at 8~GeV. The ICD-2 scenario is more cost-effective for the long-baseline neutrino program, but provides 170~kW pulsed protons at 8~GeV compared to 750~kW in the BSR scenario.  Further, both RCS-based scenarios differ from a proposed 8-GeV Linac scenario~\cite{8GL} in that the 8-GeV beam available for experiments is pulsed rather than continuous. The particle physics community should assess the required beam power and bunch format for 8~GeV beam experiments in the DUNE era in order to inform 2.4~MW planning. Similarly, the future roles of the Fermilab Recycler and Delivery Ring for the 8-GeV experimental program must also be resolved.

For all 2.4~MW RCS scenarios the reliability and performance of the proton complex may be constrained by the linac current due to activation in the H$^{-}$ injection region. Similarly any GeV-scale pulsed proton experimental program (with the PIP-II project or later) would require an accumulator ring, which would face similar challenges at injection. We propose that a 4-10~mA upgrade of PIP-II linac current be considered. Beam current and/or beam energy upgrades of the PIP-II linac need to be planned carefully with any experiments using the PIP-II beam. Presently, the Booster/RCS program and the proposed 100~kW mu2e-II program would still leaves the majority of the linac duty factor unused.

There are a number of critical areas for accelerator R\&D that could lead to substantial improvements in the performance (relative to cost) of a prospective Fermilab RCS or accumulator ring experiment program. For instance, H$^{-}$-stripping laser technology could substantially alleviate radioactivation in the injection region mentioned in the preceding paragraph. Technologies for advanced transverse optics can mitigate space-charge effects (improving beam intensity or reducing the required injection energy). New machine learning (ML) methods can be applied to optimization, diagnostics and controls to improve reliability and performance. Lastly fast-ramping superferric magnets have the potential to widen the parameter space for machine design, improve charge density, and reduce operating costs.

We call for a systematic and organized effort of the accelerator and particle physics community in defining the scientific mission of the Fermilab Proton power upgrades as well as in addressing the above critical R\&D areas.

\section{Introduction}

\subsection{Design History}

The DUNE/LBNF program has called for a 2.4~MW upgrade to achieve its ambitious particle physics experimental program. Table~\ref{tab:Milestones} shows some key DUNE/LBNF milestones with 2.4~MW power upgrade~\cite{DUNE}.

\begin{table}[htp]
  \begin{minipage}[c]{0.5\textwidth}
\includegraphics[width=1\textwidth]{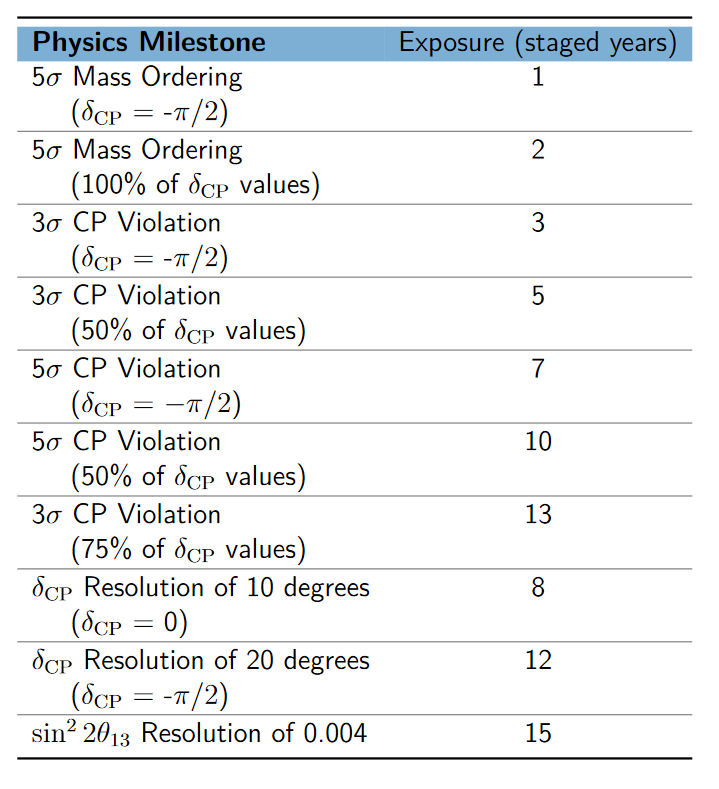}
  \end{minipage}
  \begin{minipage}[c]{0.5\textwidth}
    \caption{Progression of DUNE/LBNF physics milestones for staged scenario given in DUNE Technical Design Report. In this scenario, the 2.4~MW upgrade takes place at year six.} \label{tab:Milestones}
  \end{minipage}
\end{table}

The Fermilab Booster cannot reach the intensity (13-30~e12 protons) required to achieve 2.4~MW in the Main Injector. A recent analysis of Booster loss and emittance growth mechanisms is given in \cite{ShiltsevEldred}. After PIP-II, the geometry of the Booster will not accommodate further increases in injection energy and increasing the beam intensity may lead to prohibitive losses during transition-crossing. The Booster lacks a vacuum chamber within the main dipole magnets and the impedance from the dipole laminations are a major source of beam instabilities. The Booster is a 50-year old machine which lacks the separate-function magnets, dispersion-free acceleration, and transition-free ramp associated with a modern ring design. Further, the passive shielding of the Booster tunnel will not be adequate for high power operations and the long-term reliability of critical accelerator components cannot be guaranteed.

A scenario of achieving 2-MW beam power in the Fermilab Main Injector by replacing the Booster with a new RCS with a 2~GeV linac was originally laid out in the 2003 Proton Driver Study II (PD2)~\cite{PDriver}. In 2010, a superseding RCS proposal was provided in the Project X Initial Configuration Document 2 (ICD-2)~\cite{ICD2}. The modern 2.4~MW version of this ICD-2 scenario has provided been in \cite{Nagaitsev}; it is characterized by a 2~GeV upgrade of the PIP-II linac, a cost-effective 8-GeV RCS which ramps at 10~Hz, six batches (pulses) accumulated in the Recycler, and ultimately a 2.4~MW Main Injector at 120~GeV.

The technical challenges associated with the ICD-2 upgrade scenarios are well-considered, and achieving a 2.4~MW Main Injector for the DUNE/LBNF program is the main priority. However, variations on this scenario can be considered to provide a robust experimental program at other beam energies and opportunities for subsequent beam power upgrades.

In \cite{HarnikScience}, a community white paper was developed to survey the compelling science opportunities in next-generation of particle accelerators. In \cite{SyphersEldred}, a modification of the 2.4~MW ICD-2 upgrade scenario is described which would provide beamlines for a wide range of the experiments described in the community white paper. The goal of the design effort was not to prioritize potential experiments or restrict their siting options to Fermilab, but rather to assess the implications of incorporating these beamline requirements into the DUNE-era proton complex at Fermilab.

So now there are two similar 2.4~MW upgrade scenarios with corresponding white papers: what this document will call the modern ICD-2 scenario~\cite{Nagaitsev} and the more expansive Booster science replacement (BSR) scenario~\cite{SyphersEldred}. This document will summarize the two scenarios, compare the design features, and indicate key planning considerations. The objective is not to pit one against the other, but rather to reframe them as contingencies for each other.

\subsection{Comparison of Scenarios}

Table~\ref{tab:PDtable} gives a list of key accelerator parameters associated with the recent ICD-2 and BSR scenarios. The PIP-II era Booster accumulates 12 batches, rather than six batches, in the Recycler and Main Injector using a technique known as slip-stacking. However, losses associated with slip-stacking are likely incompatible with 2.4~MW operations (see \cite{EldredJINST,EldredNuFACT19} for an assessment of RCS scenarios if high-power slip-stacking is viable). Consequently, the ICD-2 and BSR scenarios envision a path forward with conventional boxcar stacking and beam intensities more than four times (or more) times greater than the PIP-II Booster.

We have three options for quadrupling the beam intensity in this new RCS: (1) slow down the RCS ramp rate to 10 Hz in order to have sufficient time for injection, (2) add an accumulation ring while keeping the ramp rate at 20 Hz, or (3) increase the linac beam current.

\begin{table}[htp]
\centering
\begin{tabular}{|| l | l || l || l | l ||}
\hline
Parameter & PIP-II Booster & ICD-2 & BSR v1 & BSR v2 \\
\hline
Linac Energy & 0.8~GeV & 2~GeV & 2~GeV & ~ \\
Minimum Linac Current & 2~mA & 2~mA & 2~mA & 5~mA \\
GeV-scale Accumulator Ring & Optional & Optional & Required & Optional \\
\hline
RCS Energy & 8~GeV & 8~GeV & 8~GeV & ~ \\
RCS Intensity & 6.5~e12 & 26~e12 & 37~e12 & ~ \\
RCS Circumference & 474.2~m & 553.2~m & 570~m & ~ \\
\hline
RCS Rep. Rate & 20~Hz & 10~Hz & 20~Hz & ~ \\
Number of Batches & 12 & 6 & 5 & ~ \\
Accumulation Technique & Slip-stacking & Conventional & Conventional & ~ \\
8~GeV Accumulation & Recycler & Recycler & Main Injector & ~ \\
\hline
Available RCS Power & 80~kW & 170~kW & 750~kW & ~ \\
\hline
Main Injector Intensity & 80~e12 & 156~e12 & 185~e12 & ~ \\
Main Injector Cycle Time & 1.2~s & 1.2~s & 1.4~s & ~ \\
Main Injector Power (120 GeV) & 1.2~MW & 2.4~MW & 2.4~MW & ~ \\
\hline
Upgraded Main Injector Power & ~ & 3.3~MW & 4.0~MW & ~ \\
\hline
\end{tabular}
\caption{Parameters for the PIP-II Booster as given in the PIP-II CDR~\cite{PIP2}. Parameters for the modern ICD-2 are given in \cite{Nagaitsev}, and represents a possible 2.4~MW upgrade after PIP-II. Parameters for the two BSR scenarios are given in \cite{SyphersEldred} as an alternative upgrade path to 2.4~MW. The only difference between ``v1'' and ``v2'' is the injection strategy. The Main Injector Upgrade scenario in the final row is for a 0.9~s Main Injector ramp}
\label{tab:PDtable}
\end{table}

At 2~GeV, space-charge loss mechanism are expected to be manageable even for beam intensities as high as 26~e12 or 37~e12 protons. As the accelerator design matures and the experimental program determined, an injection energy somewhat lower (or higher) than 2~GeV can also be considered.

A key difference between the ICD-2 and BSR scenario is that that the ICD-2 scenario accumulates beam in the Recycler while the Main Injector is ramping. Consequently, the ICD-2 RCS only has to ramp at a minimum of 5~Hz (in this case 10~Hz) to accumulate six batches within the 1.2~s Main Injector cycle time. The BSR scenario in contrast uses a higher ramp rate of 20~Hz, does not require using the Recycler, and extends the Main Injector cycle by 0.2~s to accumulate its five batches.

The difference in ramp rate is the primary driver of the beam power available for 8~GeV experiments (while concurrently supplying beam to the Main Injector program). The 10~Hz ICD-2 scenario provides seven 26e12 batches every 1.2~s for 8~GeV experiments (beam power up to 170~kW). While the 20~Hz BSR scenario provides twenty-three 37e12 batches every 1.4~s for the 8~GeV experiments (beam power up to 750~kW).

The combination of the 2~mA linac current and a 20~Hz ramp-rate poses a challenge. If the RCS uses a 20~Hz resonant-magnet circuit to ramp, then the main bend fields will change 1-2\% over the course of the injection time and the beam must be stabilized. One solution for the BSR scenario is to use a 2~GeV accumulator ring to facilitate injection - indicated as ``BSR v1'' in Table~\ref{tab:PDtable}. The alternative ``BSR v2'' scenario is to increase the PIP-II linac pulsed current to 5~mA in order to shorten the injection time. A 5~mA injection current would also provide an additional benefit for the ICD-2 or BSR RCS scenarios, in that it would reduce particle losses due to scattering off the injection foil.

The example BSR lattice provided in \cite{SyphersEldred} is also less compact than the example ICD-2 lattice provided in \cite{Nagaitsev}, and consequently it take five rather than six batches to fill the Main Injector. In the BSR scenario each batch must be 40\% more intense than the ICD-2 scenario to compensate for the smaller number of batches and longer cycle time. The larger circumference in the BSR is a consequence of the design choice for a highly superperiodic lattice (eight-fold symmetry, rather than two-fold) to mitigate betatron resonances. The larger circumference is also used to accommodate more RF cavities needed for the higher ramp rate (and upgradeability in ramp rate). The possibility of accommodating nonlinear integrable optics or space-charge compensating electron lens in the BSR lattice should be investigated further.

\subsection{Staging Considerations}

The RCS construction can take place concurrently with the construction for extending the energy of the PIP-II linac (and most concurrently with PIP-II Booster operations). Further, the RCS can be initially commissioned at a beam energy lower than the full 2~GeV, if necessary.

Either RCS design could likely achieve the 1.8~MW Main Injector benchmark with an injection energy of around 1.6~GeV. At 1.8~MW, the Main Injector will achieve the maximum beam power before need to upgrade Main Injector RF power (beyond PIP-II era capabilities).

The Main Injector RF upgrade may compete with the Recycler for space in the tunnel (although there are no ``show-stoppers'' identified yet). For that reason, the 1.8~MW benchmark could be a natural point to evaluate the role of the Recycler in the Main Injector program. At that point, there would also be considerable operational experience with the performance of the Recycler at intensities higher than 54e12 protons. Going to 2.4~MW, the options will be to continue to use the Recycler, to inject directly into the Main Injector, to rebuild the Recycler with improved aperture and optics, or (optimistically) to use slip-stacking in the Recycler at higher than PIP-II intensity.

However, to achieve the 2.4~MW benchmark in the Main Injector without the Recycler will require the RCS to ramp at 20~Hz (i.e. the BSR scenario). Either the RCS should be constructed with the intention of continuing to support the Recycler through 2.4~MW operations (upgrading it as necessary) or the RCS should be constructed with the capability of upgrading to a 20~Hz ramp rate.

An RCS upgradeable to 20~Hz (or more) would require metallized ceramic vacuum vessels (like J-PARC RCS) to avoid overheating the vacuum vessel with eddy currents from the main dipole and quadrupole magnets. The maximum ramp rate of the RCS may also be constrained by the number of available straights in the lattice for the installation of the RF cavities. If there is no accumulator ring to facilitate injection, the linac current should be at least 5~mA. In other words, there may an intermediate scenario that would use the ICD-2 operating model initially, but is upgradeable to the BSR operating model.

\section{Near-term \& Future Experimental Program}

\subsection{PIP-II Booster era}

In the PIP-II era~\cite{PIP2}, the Main Injector program will deliver beam to the DUNE/LBNF program as well as a fixed target slow-extraction program. At 8~GeV, the Booster can continue to supply beam to the short-baseline neutrino program~\cite{SBND} and supply beam to the new muon campus mu2e~\cite{Mu2e} program. The 2~mA CW-capable PIP-II linac will only use 1\% duty factor to supply the Fermilab Booster with its full beam power. The mu2e-II experiment~\cite{mu2e2} has been proposed to use 100~kW of the linac power, but the vast majority of the potential 1.6~MW beam power of the PIP-II linac would still be unsubscribed. The full spectrum of physics opportunities that can provided by the PIP-II linac should be pursued.

More recently, there has been a proposal for a GeV-scale hadron-absorber dark-matter experiment called PIP2BD~\cite{PIP2BD} and the possibility of hosting this experiment at Fermilab in the PIP-II era is being investigated. Ideally the PIP2BD experiment would use proton pulses 400~ns or shorter, with beam-power as high as possible and duty-factor as low as possible. A charged-lepton flavor-violation (CLFV) experiment called PRISM~\cite{PRISM_Driver,PRISM2} has also been proposed with a requirement of 30~ns pulses (one or two 50.3~MHz bunches) but to simultaneously maximize beam power. Either of these PIP-II era experiments would require a new high-power (0.1-1~MW) accumulator ring, and there is significant overlap in the design criteria for the PIP-II accumulator ring (PAR) requested by both experiments.

The high bunch charge and low pulse length favors a PAR design with minimal circumference (approximately 100~m). However an accumulator with a circumference of 474.2~m would be capable of facilitating injection into the PIP-II era Booster~\cite{BAR} and would also allow the Booster to upgrade injection energy to 1.0~GeV (with an extension of the PIP-II linac). To satisfy all three applications, two 240~m rings or four 120~m rings could be considered (similar to CERN PS Booster, except without ramping).

The prospect of a PAR beam program is mostly independent of the proposed 2.4~MW RCS upgrades. However the role of an accumulator ring for pulsed proton programs should be considered alongside any upgrades to the beam current, duty-factor, and/or energy of the PIP-II linac. Like the RCS program, foil-injection into the PAR beam program also benefits from increasing the 2~mA PIP-II beam current to 5~mA or more. The PAR beam program can be designed for a particular beam energy, or could be designed to increase in energy when the energy of the PIP-II linac is extended.

\subsection{2~GeV Linac \& RCS}

The experiments listed in \cite{HarnikScience} are organized in \cite{SyphersEldred} according to which proton energy and pulse structure is required (except those experiments requesting electrons or positrons). There are experiments using high-duty factor 2~GeV beam from the upgraded PIP-II linac, 2~GeV pulsed protons from a possible accumulator ring, 8~GeV pulsed protons directly from the RCS, DUNE/LBNF beamline operating 60-120 GeV from the Main Injector, and 120~GeV beam slow-extracted from the Main Injector.

In the present configuration, the muon g-2 and mu2e experiment use the Recycler and Delivery Ring to manipulate the muon/proton beams for the experiments (momentum selected-muons and slow-extracted protons). The operational years of the mu2e experiment might overlap with the operational years of the RCS beam program. As the rest of the 8-GeV experimental program is evaluated for the RCS-era, the role and required performance for the Recycler must be determined.

The ICD-2 and BSR scenario differ in their delivered beam power for the 8-GeV beam program, however there is some convergence in the available beamlines. While the 2-GeV accumulator ring provides greater operational benefit to the BSR RCS design than the ICD-2 RCS design, either scenario can feature an accumulator ring for the experimental program. Similarly, although the ICD-D scenario requires use of the Recycler for stacking beam into the Main Injector, the BSR scenario could also maintain the Recycler for the experimental program.

\subsection{Future Upgradeability}

The Linac and RCS upgrade is not just a means to achieve 2.4~MW for the DUNE/LBNF program, but will shape the architecture of the Fermilab proton complex for decades thereafter. Accordingly, the opportunities for subsequent upgrades should be identified.

The performance of the RCS might not gain any further benefit by raising the injection energy above 2~GeV, as space-charge effects may no longer be the dominant driver of injection losses. After the RCS injection point, however, the PIP-II linac can be extended beyond 2~GeV to deliver to any experiments which benefit. The original Project-X design called for a 3~GeV experimental program~\cite{ICD2}.

At 8~GeV, beam power is driven primarily by the RCS ramp rate. In the BSR scenario, a potential 30~Hz upgrade is described as a route to a 1.2~MW 8~GeV program. The maximum upgradeable ramp-rate of RCS needs to be incorporated as a design requirement before the RCS is built, which in turn is an evaluation of the future 8~GeV program.

At 120~GeV energy, the most reliable upgrade would be to increase the ramp-rate of the Main Injector itself. With the 0.9~s cycle time outlined in \cite{SyphersEldred}, the BSR scenario would achieve 4~MW or the ICD-2 scenario would achieve 3.3~MW. The DUNE/LBNF target hall is designed for 2.4~MW, consequently a 3-5~MW long-baseline neutrino program would likely require constructing a second target station.

\section{Beneficial R\&D}

This section will cover some key areas for accelerator R\&D including H$^{-}$-stripping laser systems, optics for mitigating space-charge effects, ML-enhanced diagnostics and controls, and fast-ramping superferric magnets.

\subsection{Laser H- Injection System}

Currently the H$^{-}$ linac beam can be accumulated in a high-intensity proton ring via charge-stripping foil injection, in which a thin carbon-based foil target is used to remove the two electrons from the H$^{-}$ beam and combine it with a circulating proton beam.

At both the Oak Ridge SNS ring and the J-PARC RCS ring, the overwhelmingly highest radioactivation levels are found at the injection region and the area immediately downstream from injection. This activation is driven primarily by injection protons scattering off the injection foil in the duration of time while the linac fills the ring. The foil itself heats to 1600-2200~K as a result of beam interaction, and consequently foil lifetime becomes a major concern for machine reliability. With present technology the foil injection process is a major design constraint on the Fermilab RCS, as well as any GeV-scale accumulator ring for pulsed proton experiment (see discussion in \cite{EldredJINST}).

H$^{-}$ beams can also be stripped by the combination of high-power lasers and strong magnetic fields, recently demonstrated at the SNS over a 10~$\mu$s timescale with 95\% stripping efficiency~\cite{Cousineau}. However, further development is needed to extend this technique over a multi-ms timescale, provide a greater stripping efficiency, and maintain efficient use of laser power. In the meantime, RCS and accumulator ring upgrade scenarios can consider laser-injection schemes as an alternate or retrofit of the foil-injection baseline.

\subsection{Mitigation of Space-charge}

For a fixed injection energy and machine aperture, the intensity of an RCS or accumulator ring will be limited by the maximum obtainable incoherent space-charge tune-spread $\Delta Q_{SC}$. An empirical model for the relation between space-charge tune-spread, emittance-growth, particle loss, and beam intensity is given in \cite{ShiltsevEldred}. Any improvement to the $\Delta Q_{SC}$ loss limit provides immediate value for particle accelerator performance (relative to project cost).

Transverse and longitudinal painting can be used to limit $\Delta Q_{SC}$ and high-efficiency collimators can be used to managed particle losses. Several betatron resonances have to be corrected simultaneously to increase the maximum obtainable $\Delta Q_{SC}$. For an RCS or accumulator with transverse optics in $n$ perfectly repeating periods (superperiodicity), the particle losses in theory scales with $\Delta Q_{SC}/n$ instead. The full advantage of machine superperiodicity is rarely obtained in reality, and consequently the extent to which a future RCS or accumulator ring can benefit from a superperiodic lattice is a valuable design question. The Fermilab Booster, which currently features a unique combination of extreme space-charge and near 24-fold superperiodicity, might serve as a useful source of space-charge R\&D.

The Integrable Optics Test Accelerator (IOTA) and Fermilab Accelerator Science \& Technology (FAST) facility were created for advanced accelerator R\&D and beam physics research~\cite{AntipovIOTA}. Nonlinear integrable optics and electron lens space-charge compensation are two additional technologies under development at FAST/IOTA to be tested for their application to intense hadron rings in the coming years.

The electron lens is a versatile particle accelerator device with applications in beam-beam compensation, collimation, nonlinear focusing, and space-charge compensation. In \cite{SternShiltsev}, particle loss and emittance growth under extreme space-charge is simulated, and the mitigation of emittance growth by a series of electron lenses is demonstrated. Further work is needed in experiment and simulation to understand the ultimate potential and correct application of this technology.

Nonlinear integrable optics is an innovation in acceleration design to provide immense nonlinear focusing without generating parametric resonances~\cite{DanilovNagaitsev}. Strong nonlinear focusing could significantly enhance the performance of an intense hadron ring by mitigating halo formation and damping collective instabilities. In \cite{EldredValishev}, space-charge and chromatic effects in an example RCS are simulated and the mitigation of halo formation is demonstrated.

\subsection{Machine Learning}

Compared to traditional data science tools, machine learning (ML) methods allow (i) complex or distinct types of data to be synthesized; (ii) many iterations of data including those with rare phenomena to be aggregated; and/or (iii) relationships between the data to be derived empirically, independent of prior modeling. In \cite{EdelenICFA} an overview of specific ML applications for particle accelerators is given - anomaly detection, online modeling, surrogate modeling, virtual diagnostics, online tuning, and data analysis.

At this critical juncture of RCS and accumulator ring design, particle accelerator simulation and lattice optimization should also take advantage of modern ML techniques to achieve more rapid and comprehensive results. In addition, ML techniques have demonstrated the potential to improve the performance, interpretability, and reliability of present-day machines.

Although ML techniques have found their earliest success in precision electron machines, extending these techniques to intense hadron rings could offer powerful gains in performance. ML-enhanced tuning can facilitate resonance-correction and control of local optics in intense hadron rings. Anomaly detection could improve identification of beam instabilities and components failure. ML techniques have begun to be applied to improve the accuracy of profile monitor diagnostics and to separate loss mechanism recorded by beam loss monitors. The present generation of intense hadron rings could serve as a valuable test-bed for next-generation ML controls and diagnostics.

Some challenges for applying ML techniques to particle accelerator controls and diagnostics include - operational restrictions on parameter tuning, undocumented/unmeasured changes, and sparse data collection relative to the large parameter space. Consequently these ML applications will benefit from the close collaboration between individuals with expertise in algorithms and those with accelerator domain knowledge.

\subsection{Fast-ramping Superferric Magnets}

Recently a 300 T/s dipole ramp rate was demonstrated in a REBCO-based high-temperature superconductor superferric magnet~\cite{Piekarz}, which far exceeds the dipole ramp rate requirements of a fast RCS. The example dipole magnet design applies the dipole field to two vacuum chambers simultaneously to make full use of the return flux. With the superferric magnets, the electrical cost of cryogenic cooling is small than the electrical cost of powering the resistive coils (for a net gain).

Fast-ramping superferric magnets have the potential to greatly improve on the field strength and aperture offered by conventional resistive magnets, which could have transformative implications for RCS design.

With a smaller circumference (for the same extraction energy and beampipe aperture) each RCS batch would have a comparable intensity but a higher linear charge density. The smaller circumference would also improve the acceleration efficiency of RF systems (as each RF cavity provides acceleration per turn inversely proportional to revolution period). 

A larger RCS magnet aperture could allow for a beam with larger emittance and consequently greater intensity relative to $\Delta Q_{SC}$ (or lower injection energy). The beam emittance may ultimately be limited by the Recycler and Main Injector acceptance, but the 8~GeV program would not be so limited.

High-field, fast-ramping super-ferric magnets open up a larger machine-design parameter space than previously considered in the ICD-2 or BSR scenarios described above, and further design work would be needed to optimize a superferric RCS upgrade scenario.

\bibliographystyle{JHEP}
\bibliography{references}  

\end{document}